\documentclass[fleqn,10pt]{wlscirep}
\usepackage[utf8]{inputenc}
\usepackage[T1]{fontenc}
\usepackage{xcolor}
\usepackage{amssymb}

\title{Sculpting ultrafast mid-infrared light for solid-state high-harmonic generation}

\author[1,*]{Camilo Granados}
\author[2]{Bálint Kiss}
\author[2,3]{Eric Cormier}
\author[4]{Bikash Kumar Das}
\author[2]{Debobrata Rajak}
\author[5]{Carmelo Rosales-Guzman}
\author[2]{Rajaram Shrestha}
\author[4,6,7]{Stephan Fritzsche}
\author[8,9,10]{Qiwen Zhan}
\author[1,*]{Wenlong Gao}

\affil[1]{Eastern Institute of Technology, Ningbo 315200, China}
\affil[2]{ELI-ALPS, ELI-Hu Non-Profit Ltd., Szeged, Hungary.}
\affil[3]{Laboratoire Photonique Numérique et Nanosciences (LP2N), UMR 5298, CNRS-IOGS-Université
Bordeaux, 33400 Talence, France}
\affil[4]{Theoretisch-Physikalisches Institut, Friedrich-Schiller-Universität Jena, Max-Wien-Platz 1, D-07743 Jena, Germany}
\affil[5]{Centro de Investigaciones en Óptica, A.C., Loma del Bosque 115, Colonia Lomas del campestre, 371507 León, Gto., Mexico.}
\affil[6]{Helmholtz-Institut Jena, D-07743 Jena, Germany}
\affil[7]{GSI Helmholtzzentrum f\"ur  Schwerionenforschung GmbH, D-64291 Darmstadt, Germany}
\affil[8]{School of Optical-Electrical and Computer Engineering, University of Shanghai for Science and Technology, Shanghai 200093, China}
\affil[9]{Zhejiang Key Laboratory of 3D Micro/Nano Fabrication and Characterization, Department of Electronic and Information Engineering, School of Engineering, Westlake University, Hangzhou 310030, China}
\affil[10]{International Institute for Sustainability with Knotted Chiral Meta Matter (WPI-SKCM2), Hiroshima University, Higashihiroshima, Hiroshima 739-8526, Japan}
\affil[*]{wgao@eitech.edu.cn}
\affil[*]{cagrabu@eitech.edu.cn}

\begin{abstract}
The ability to sculpt light in space, time, and polarization has revolutionized studies of light–matter interaction and enabled breakthroughs in optical communication, imaging, and ultrafast science. Among the many degrees of freedom of light, orbital angular momentum (OAM) further expands these capabilities by unlocking new regimes of control in information encoding, particle trapping and manipulation, and symmetry-driven selection rules. However, exploiting OAM to drive nonlinear, non-perturbative effects in solids remains challenging, especially in the mid-infrared (MIR) spectral regime-a key region for accessing these effects in ambient air, where spatial light modulators do not operate. Here, we circumvent this limitation by generating femtosecond, few-cycle MIR Bessel-Gauss vortex (BGV) and perfect optical vortices (POVs), using a robust, static spatial-shaping strategy. By utilizing these beams to drive nonlinear optical processes such as second-harmonic generation (SHG) and high-harmonic generation (HHG) in various solid-state materials, we show that the resulting harmonic beams faithfully inherit the structural characteristics of the drivers: the constant-intensity ring of the POVs is preserved across harmonic orders, while the BGV harmonic beams retain their intrinsic topological charge-dependent intensity profiles. Furthermore, by verifying the linear OAM up-scaling law, we confirm the conservation of OAM during SHG and HHG in solids. These results establish strong-field HHG in solids as a robust platform for synthesizing ultrafast structured harmonic light with controllable, high-value OAM. 
\end{abstract}

\begin{document}

\flushbottom
\maketitle
%
%

\thispagestyle{empty}


\section*{Introduction}

Light beams carrying orbital angular momentum (OAM) \cite{Allen_OAM}- also known as vortex beams (VBs)- have profoundly reshaped our understanding of light–matter interactions across various scales \cite{Anomalous,NonLinear_Compton,Nuclear_eVortex,elec_posit, TransOAM, Forbes}. In both perturbative and non-perturbative optical regimes, these interactions are typically described within the framework of the dipole approximation \cite{Laura}, yet recent advances have clearly shown effects beyond the dipole limit \cite{Helical_dichro,Helical_dichro2,Helical_dichro3}. However, exploring how matter responds to different properties of VBs, particularly to their OAM (or, their topological charge (TC)), is often constrained to low OAM values. This limitation typically arises from the OAM conservation law, which leads to an enhancement in the beam size with increasing values of the TC \cite{Divergence_OAM}, hindering high-intensity-dependent nonlinear processes such as high-order harmonic generation (HHG), where maintaining a sufficient peak intensity is essential to surpass the nonlinear threshold \cite{HHG_Theory}. 

A robust solution to this limitation is the utilization of perfect optical vortex (POV) beams. Unlike conventional vortex beams such as Laguerre-Gaussian and Bessel-Gauss beams, POV beams are characterized by a ring width that is smaller than their ring radius as well as an annular intensity distribution which remains invariant with the TC \cite{POV}. This invariant geometry of the beam intensity is essential in experiments where high spatial resolution is crucial \cite{STED_vortex}, as well as in optical encryption, optical communication, optical tweezing, and microscopy, among others~\cite{Vanitha_2021,YangXieZhang,SHAO2018545,Chen:13,Chonglei}. However, realizing POV beams in the femtosecond (fs) temporal and long-wavelength regime, which are uniquely suited for investigating nonlinear non-perturbative processes in solid-state systems, presents a threefold challenge: (1) the scarcity of specialized ultrashort mid-infrared (MIR) laser sources, (2) the absence of dynamic beam-shaping optics (e.g., spatial light modulators) in the MIR spectral regime, and (3) the difficulty of preserving femtosecond pulse durations during complex spatial shaping. 

The generation of MIR structured light based on static elements is then pivotal to overcome these limitations and to advance our knowledge on non-perturbative effects in matter. As such, the ultrashort structured light reported in this manuscript provides a versatile platform to investigate light-matter interactions in solid-state systems which intrinsically exhibit diverse symmetries, topologies, and spin–orbit couplings. As a result, ultrashort structured light opens the door to address fundamental questions concerning OAM conservation in solid-state HHG. Additionally, ultrashort structured light driven HHG establishes the harmonic process in solids as a source of coherent, compact, and short-wavelength exotic vortex beams as the POV and BGV beams in ambient air. Consequently, the requirement of large vacuum systems which are typically employed in gas-phase HHG experiments can be bypassed. These developments created new opportunities to investigate quantum entanglement in solids and to produce structured quantum states of light with a large photon number for tailoring light-matter interactions \cite{ExtremeOAM}. Furthermore, MIR structured light driving the solid-state HHG is a potential source of twisted attosecond pulses \cite{AttoTwisted} in the vacuum-ultraviolet (VUV) spectral range \cite{Atto1}, and of exotic topological structures such as skyrmions and Hopfions. 

In this work, we implement a completely passive beam-shaping strategy specifically designed for the MIR domain, which preserves the ultrashort temporal structure but still enables us to structure the light spatially. Our approach combines a spiral phase plate (SPP), an axicon (AX), and a lens \cite{POV_FT, Artl2000} (Fig.\ref{Fig1} top). With these optical elements in hand, we demonstrate the generation of MIR (3.2~$\mu$m) spatial POV and Bessel–Gauss vortex (BGV) beams in the fs regime with few-cycle pulses suitable for strong-field physics experiments. We verify the TC-invariant features of the fundamental POV beams by generating them with multiple TCs, showing that their ring radius and annular intensity distribution remain independent of the TC. This behaviour contrast to the TC-dependent transverse intensity distribution exhibited by BGV beams. Using the fs fundamental POV beam, we first investigate the frequency-doubling process in zinc oxide (ZnO; a 3D, direct, wide-band gap semiconductor typically used for scalable UV devices) and gallium selenide (GaSe; a layered van-der-Waals semiconductor suitable for 2D optoelectronics). We then extend these measurements to high-harmonic generation (HHG) in the same materials. The selection of these two materials provides a direct comparative study of how the OAM of the driver couples to materials exhibiting different crystal symmetries and dimensionalities. Furthermore, we study the conservation law of the OAM for these vortex beams and show the robustness of the HHG process in solids as a source of high TC structured light. These experiments demonstrate that the fs POV pulses effectively drive both perturbative and non-perturbative nonlinear processes, producing short structured pulses across the infrared-to-visible spectral range with large values of the TC.

\section*{Results}
The fundamental POV beam can be described in two distinct regimes, as characterized by the ratio between the beam radius (or ring radius), $r_0$, and the half-ring width, $\delta$. In the regime, where $r_0/\delta>1$, though not too high, the complex field amplitude of the POV beam is mathematically expressed as (at the source plane): 

\begin{equation}
    \tilde{E}(r,\theta)=E_0\exp{\left(-\frac{(r^2+r_0^2)}{\delta^2}\right)}\exp{(il_0\theta)}I_{l_0}\left(\frac{2rr_0}{\delta^2}\right), \label{eq1}
\end{equation} 

where $l_0$ is the TC carried by the beam, ($r,\theta$) denote the polar coordinates of the beam with $r^2 = x^2 + y^2$ and $\theta = \arctan(y/x)$, $I_{l_0}(...)$ is the modified-Bessel function of the first kind of order $l_{0}$, and $E_{0}$ is the constant field amplitude. In the asymptotic regime, where $r_0 / \delta \gg 1$, in contrast, the modified-Bessel function of the first kind can be approximated to an exponentially growing function, independent of the TC, as $I_{l_0}(2r_0r/\delta^2) \approx \exp(2r_0r/\delta^2)$. In this regime, the POV field distribution takes the form: 

\begin{equation}
  \tilde{E}(r,\theta)=E_0\exp{\left(-\frac{(r-r_0)^2}{\delta^2}\right)} \exp{(il_0\theta)}.  \label{eq2}
\end{equation}

The asymptotic regime is reached when the beam radius, $r_0$, exceeds the beam half-ring width, $\delta$, by approximately an order of magnitude or more~\cite{HHG_POV}. As follows from Eq.~\ref{eq2}, the field amplitude is entirely independent of the TC, leading to a TC-independent intensity distribution for POV beams characterized by a bright ring at $r=r_{0}$. In contrast, Bessel-Gauss vortex (BGV) beams \cite{BG_Review}, typically characterized by a TC-dependent intensity distribution and the presence of multiple concentric off-axis rings around the central vortex core, are described by the presence of a Bessel function of the first kind, $J_{l_{0}}(k_r \rho)$, and a Gaussian envelope in their complex spatial amplitudes: 

\begin{equation}
E(\rho,\phi)
= E_{0}^{'}\,
\exp\!\left(-\frac{\rho^{2}}{w^{2}_0}\right)
J_{l_{0}}\!\left( k_r \rho \right)\exp(il_{0}\phi). \label{BGV}
\end{equation}

Here, $k_r$ is the radial component (responsible for controlling the beam size) of the total wave number, $k=\sqrt{k_{r}^{2}+k_{z}^2}=2\pi/\lambda$, with $k_{z}$ and $\lambda$ representing the longitudinal wave number and the wavelength of the beam, respectively, $(\rho,\phi)$ are the polar coordinates of the BGV beam, and $E_{0}'$ is a constant amplitude. More importantly, the POV and BGV beams are Fourier pairs, i.e., $\tilde{E} (r, \theta ) \propto \mathcal{F} \left[E(\rho, \phi)\right]$. This characteristic make it possible to generate either of them by exploiting the optical Fourier transformation (FT) as will be shown below.

\begin{figure}[h!]
\centering 
\includegraphics[width=1\linewidth]{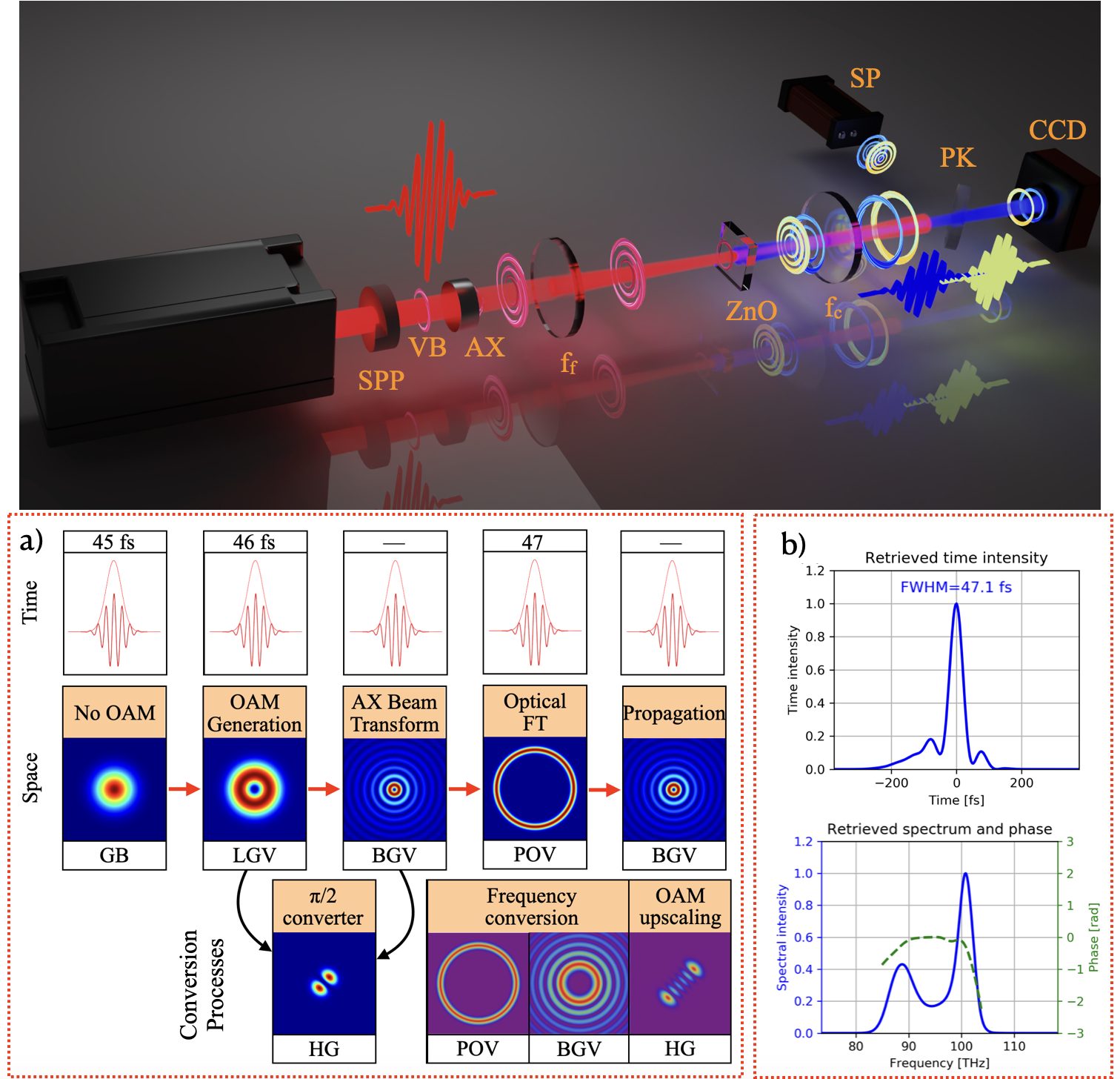}
\caption{Simplified experimental setup for the generation and characterization of fs BGV and POV beams. Here, we illustrate the generation of a fs Gaussian beam, centered at $\lambda=$ 3.2~$\mu$m, followed by the generation of the LGV and BGV beams by means of a spiral phase plate (SPP) and an axicon (AX), respectively. The generation is followed by the characterization of both fundamental and POV harmonic beams. (a) conceptualization of the experimental setup that realized the theoretical ideas supporting the experiment design. (d) Resulting  spectrum and temporal intensity extracted from the FROG traces \cite{frog} showing the temporal duration of the pulse after the axicon optical element. SPP: Spiral phase plate, VB: Vortex beam, PK: pickup mirror, $f_f$: focusing lens, $f_c$: collimation lens, SP: spectrometer and CCD: beam profiler.}
\label{Fig1}
\end{figure}

In Fig.~\ref{Fig1} (Top), we show a schematic representation of the experimental setup at the Extreme Light Infrastructure - Attosecond Laser Pulse Source (ELI-ALPS, MIR laser source) used to generate fundamental VBs and their corresponding harmonic vortices via the SHG and HHG processes. The experimental setup comprises of two main sections summarized as follows: The first section deals with short pulses generation, $\tau_{p}\approx 45$~fs (4 cycles, average duration over the beam aperture) with $\tau_{p}$ being the pulse duration, in the form of Gaussian beams centered at a wavelength of $\lambda=3.2$~$\mu$m with a repetition rate of 100~kHz (delivered by the MIR laser system)~\cite{Laser_System}. The energy per pulse of the fundamental beam is, $E_{p}\approx 140$~$\mu$J (corresponds to a peak intensity of $I_{\text{peak}}=1.6\times 10^{12}$ W/cm$^2$). The fundamental Gaussian beam is then guided towards the spiral phase plate (SPP, Vortex Photonics, Munich), which imparts the required helical phase onto the Gaussian beam and transforms it into fs Laguerre-Gaussian vortex (LGV) beam. When the LGV beam passes through an axicon (AX, Vortex Photonics, Munich) optical element, fs BGV beam is produced. Both LGV and BGV beams, characterized with a MIR camera (DIAS Infrared systems), share the same T. 

The second section of the MIR laser source setup deals with the characterization of the fundamental POV and BGV beams, and the harmonics generated when these beams drive the harmonic generation process in various solids. We adopt the following strategy to generate fs POV beams from fs BGV beams: we use a collimating lens ($f_c=125$~mm) at a distance of $\approx 35$~cm from the AX element and a focusing lens ($f_f=75$~mm) at a distance of $\approx 45$~cm from the collimating lens, to generate fs POV beams at its focal plane. The focused beam is then imaged into an imaging camera (Basler acA1440, spectral range 1200–300~nm). For simplicity, in Fig.~\ref{Fig1} (Top), we call this camera a charge-coupled device (CCD). Upon propagation, fs POV beams transform into fs BGV beams at a distance of $\approx 2f_f$ from the focusing lens, because their Fourier pairs nature, which can be exploit via their propagation to the far-field or via the optical FT. To generate harmonic vortices, we place the solid crystal between the two lenses and in the optical path of the POV beam, as exemplified in Fig.\ref{Fig1} (Top) by the ZnO crystal. Additionally, we interchange the positions of the $f_c$ and $f_f$ lenses to achieve a tighter focusing into the target materials, thereby, enhancing the harmonic generation process. 

Furthermore, we measure the TC of the fundamental and harmonic vortex beams by guiding a part of those beams towards a cylindrical lens (CL), which typically works on the principle of mode conversion i.e., the CL transforms the input OAM modes into tilted Hermite-Gaussian (HG) modes \cite{Allen_OAM,Beijersbergen1993,Padgett2002}. The number of intensity minima, $n$, between the bright lobes in the resulting tilted HG intensity distribution is directly linked to the TC of the beam as, $l_0=n$. This measurement technique can also be implemented with a tilted convex lens, where a convex lens tilted in one of the transverse dimensions transforms an input OAM mode into a tilted HG pattern at a specific distance after the lens \cite{OAMtec,HG2}. The conceptual ideas explaining the experimental design to generate fs MIR structured light beams are presented in Fig.~\ref{Fig1}(a). We display the simulated transverse spatial intensity profiles of the fundamental Gaussian, LGV, BGV, and POV beams along with their corresponding temporal durations measured during the experiment. Additionally, we show the transformed HG intensity patterns resulting from the passage of VBs through the CL. Moreover, we show spatial intensity structures of the frequency up-converted POV and BGV beams along with their TC measurement results. From the perspective of fundamental beams, two characteristics are more important: (1) their spatial integrity, and (2) their temporal duration. The latter characteristic can be ensured by using specialized optics designed for the MIR spectral regime that minimize the temporal stretching of the pulse and by compensating the residual temporal chirp. This guarantees an optimal pulse duration for driving the HHG process. 

Moreover, we can manipulate the spatial structure of the beam by leveraging on the beam transformation: to generate BGV beams, we illuminate the AX element with LGV beams. Later, the optical FT of BGV beams results in POV beams. The mode and frequency conversion of the beams are used to extract the TC of the fundamental beams, to generate ultrashort harmonic vortex beams, and to test the conservation law of OAM in solid-state HHG. Finally, we measure the fs temporal duration of the generated beams in a frequency-resolved optical gating (FROG) setup \cite{frog}. As shown in Fig.~\ref{Fig1}(b), the retrieved time intensity profile shows a pulse duration of $\approx$47.1~fs, measured after the AX optical element. Besides the time intensity, we also display the retrieved spectral intensity and phase of the driving field. To extract the temporal duration of the pulse, we mimic the effect of different optical elements with a 4~mm thick Yttrium aluminum garnet (YAG) piece. This corresponds to measure the pulse's temporal duration after $f_c$. The short pulse duration is achieved by compensating the dispersion induced by different optical elements with the Dazzler CEP control module of the MIR laser. We perform the CEP optimization by maximizing ethe HHG signal. 

\subsection*{Fundamental ultrashort MIR vortex beams}

\begin{figure}[ht!]
\centering 
\includegraphics[width=1\linewidth]{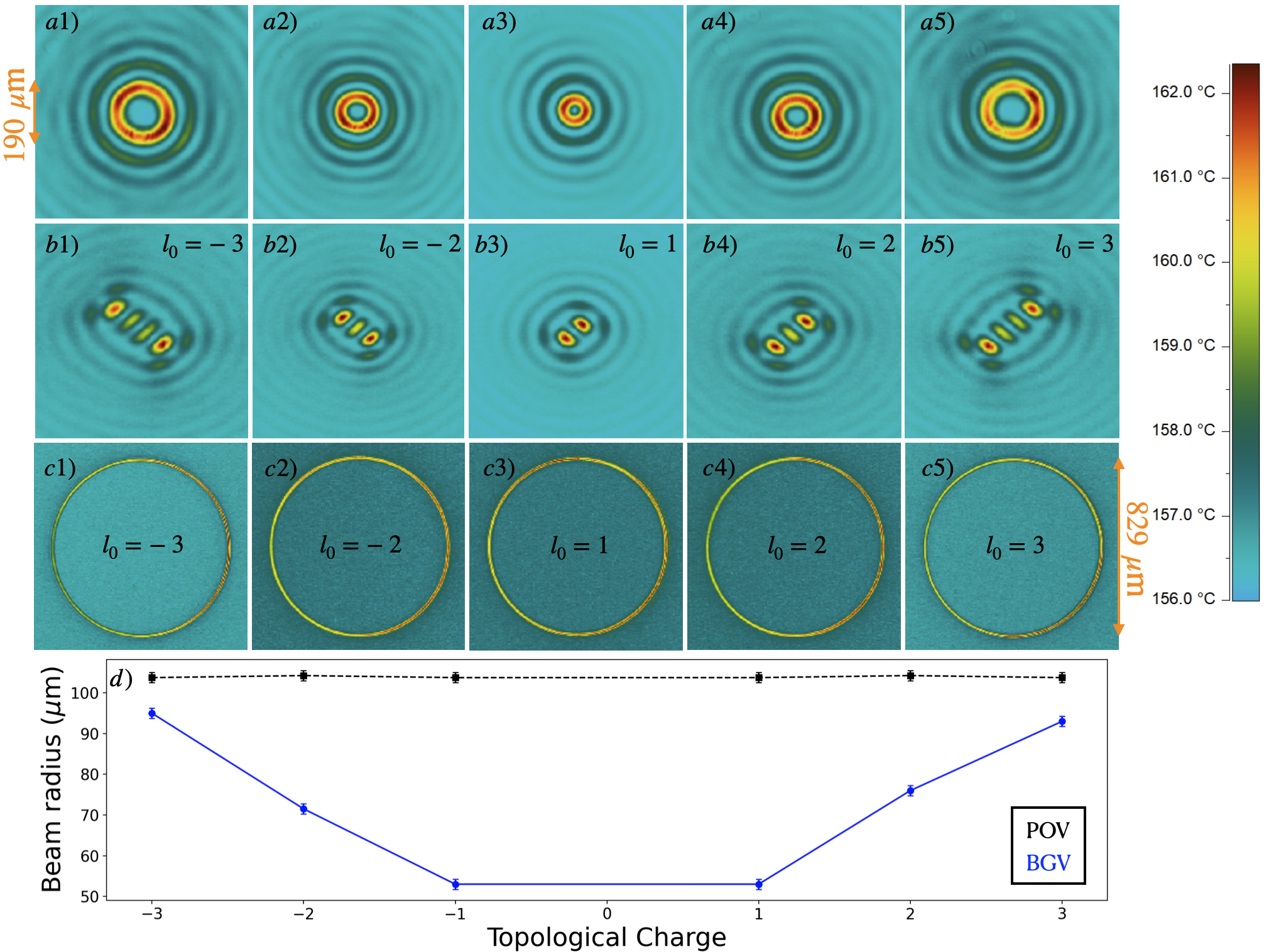}
\caption{Intensity distribution of the fs MIR BGV and POV beams. (a1)-(a5) Intensity distribution of fs BGV beams with values of the TC $l_0=-3,-2$,1,2 and 3. (b1)-(b5) Their corresponding HG intensity distribution after the transformation by a CL, from where we extracted the beam's TC. (c1)-(c5) Intensity distribution of fs POV beams generated from the BGV beams shown in (a1)-(a5). Notice that we used the same beam size value for the BGV beam generated with $l_0=1$ and $l_0=-1$ for completeness, since the latter values were not measured. (d), Comparison between the POV and BGV beam size results for the different TC. Notice that we divided the POV beam radius values by 4 for a better comparison with the BGV beam. The intensity bar on the right side corresponds to the measured temperature produced by the beam in the MIR camera. }
\label{Fig2}
\end{figure}

Figure.~\ref{Fig2} (a1)-(a5) demonstrate the generation of fs MIR BGV and POV beams, where we display the TC-dependent transverse intensity distribution of BGV beams, measured at the plane $2f_f$. It is clear from the figure that the radius of the maximum intensity of BGV beams grows monotonically as the TC increases- a characteristic also observed in LGV beams case. WE note that irrespective of the nature (continuous-wave or pulsed) of the BGV beam, its beam size expands with the TC. Additionally, in Figs.~\ref{Fig2} (b1)-(b5), we show their transformed intensity patterns after the BGV beams has passed through a CL. This pattern also shows how a CL transforms the multi-ring intensity patterns of a BGV beam into tilted HG intensity distribution with surrounding rings. As discussed earlier, the number of minima, $n$, in the transformed intensity distribution of a BGV beam is directly tied to the magnitude of its TC, $l_0=n$, whereas the orientation of the tilted lobed-intensity patterns is linked to the sign of the TC: right-tilted patterns correspond to the positive sign, whereas, left-tilted patterns to the negative sign \cite{OAMtec}. Therefore, the beam's TC shown in (b1)-(b2) are negative, while the values in (b3)-(b5) are positive. Furthermore, in Figs.~\ref{Fig2} (c1)-(c5), we show transverse intensity distribution of POV beams, measured at the plane $f_f$. We observe from Figs.~\ref{Fig2} (c1)-(c5) that the ring radius and ring thickness of POV beams hardly change with increasing TC as required for a “perfect” beam. Although this feature has been demonstrated previously with continuous-wave type POV beams, we translate such a novel characteristic to the fs temporal regime for the first time. Note that the focusing lens transforms the fs BGV beam into a POV beam, as a consequence of their conjugate FT relationship~\cite{POV_FT}.

To make the “perfectness” of the fs POV beams and show the enhancement of fs BGV beam's size with increasing TC explicit, we compute the ratio between the ring diameter and ring width for POV ($2r_{0}/2\delta$) and BGV ($2\rho_{BG}/2w_{0}$) beams, where $2\rho_{BG}$ and $2w_0$ denote the diameter and width of the maximum intensity ring of BGV beams. We note that extracting the beam diameter is easier straight forward and leads to the same conclusion than using the beam radius. Additionally, for all quantitative aspects of the fs BGV beams, here we consider the maximum intensity ring only. This description follows from the definition of divergence both in the near- and far-field, which is typically given in terms of the maximum intensity ring. For the case of fs BGV beams, shown in Figs.~\ref{Fig2}(a1)-(a5), we obtain the ratio, $2\rho_{BG}/2w_0=$ {(4.1(5), 3.0(5), 2.8(5), 3.3(5), 4(0.5)))}, for TC $l_0=-3,-2$, 1, 2, and $3$, respectively. The error assigned to the measured ring diameter and ring width corresponds to the pixel size, $5~\mu$m. Moreover, the ring diameter values are extracted to be $2\rho_{BG}=$(190(5),143(5), 106(5),152(5),186(5))$\mu$m, exhibiting a clear larger value for increasing TCs. On the other hand, for POV beams (shown in Figs.~\ref{Fig2} (c1)-(c5)), we obtain $2r_0/2\delta=$(48.5(5), 48.8(5), 48.8(5), 49.3(5), 48.8(5)), for TC $l_0=-3,-2$,1,2, and $3$, respectively. The small deviations observed in the ratio, $2r_0/2\delta$ (still similar within the error bar), for POV beams mainly originate from minor variations in the ring diameter, $2r_0=(415(5),412(5),415(5),419(5),415(5)) \mu$m, since the ring width remains constant for all the cases presented here i.e., $2\delta=$17(5)$\mu$m. Notice that the error assigned to the experimental values corresponds to the minimum pixel size, which is 5$\mu$m. In Fig.~\ref{Fig2}(d), we further compare the radius of the maximum intensity of BGV beams (blue solid line) and POV beams (black dotted line) for different TCs. For a better comparison, we divide the radius of POV beams by four. The TC-independent (dependent) transverse intensity distribution of the POV (BGV) beams is clearly demonstrated in the figure: a flat (or, constant) line for POV beam's radius with increasing values of the TC and a linear relationship between the beam radius and the TC for BGV beams. Additionally, these results confirm that the intensity distribution of generated POV beams are well-described by their asymptotic form, as displayed in Eq.~\ref{eq2}.

\subsection*{Second-harmonic generation with fs POV }

\begin{figure}[ht!]
\centering 
\includegraphics[width=1\linewidth]{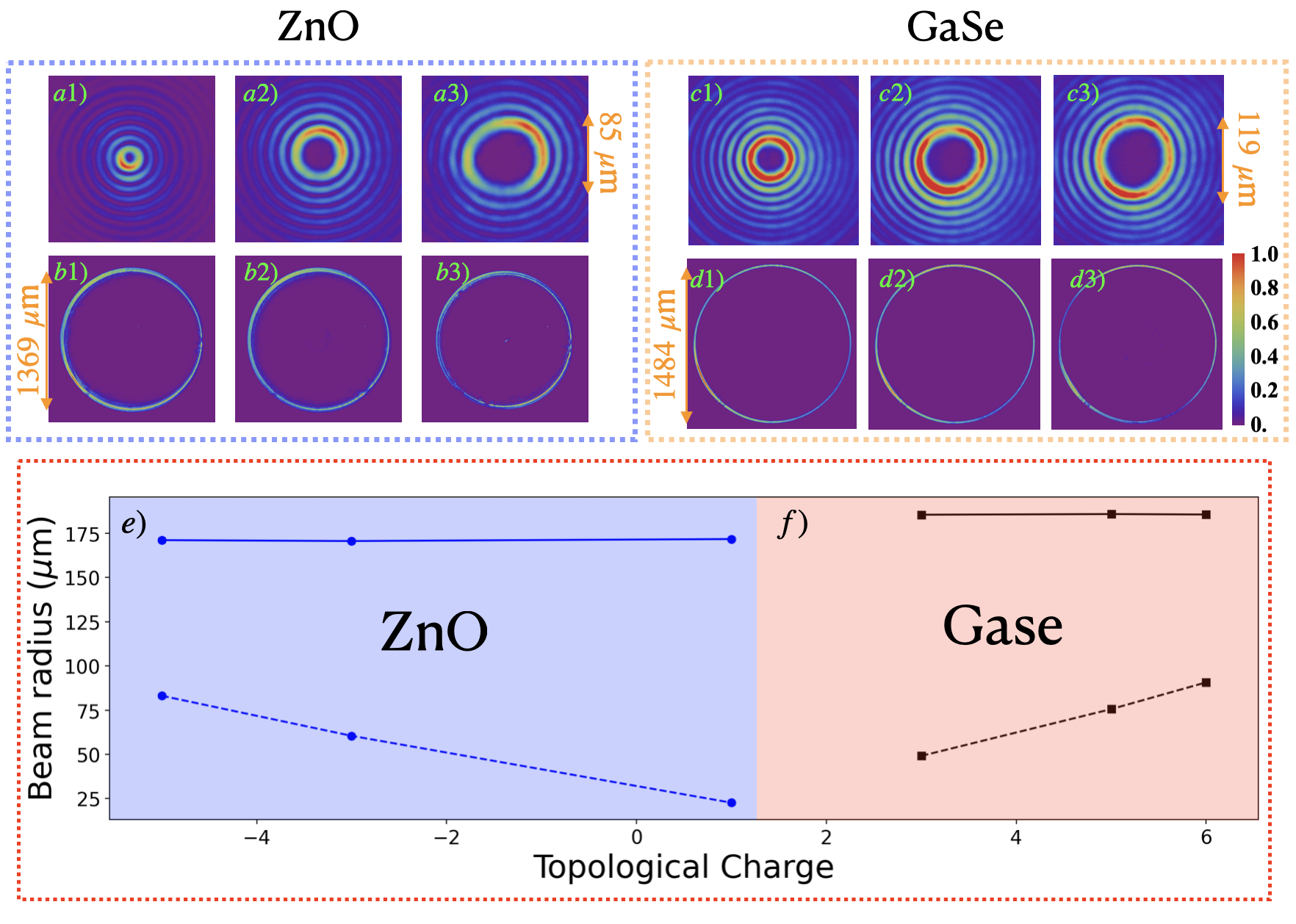}
\caption{Second-harmonic generation in ZnO and GaSe for fundamental POV and BGV (after background subtraction). (a1)-(a3) and (b1)-(b3) Normalized intensity distribution of the SH BGV and POV beams for a ZnO target, respectively. The driving field carried $l_0=1,-3$, and -5. (c) and (d), SH in GaSe for a fundamental field with values of the TC $l_0=3,5$ and 6. (e) and (f) Summary of the results for ZnO and GaSe, showing the SH POV beam size (full lines) and the SH BGV beam size (dashed lines). We divided the beam radius of the SH POV by 4 to make easily to compare the sizes of the vortex beams. The error bar for all the values corresponds to 5~$\mu$m. }
\label{Fig5}
\end{figure}

To make use of the MIR fs POV beams, we first investigate the perturbative nonlinear process of second-harmonic generation from different solid-state materials. In particular, we produce the second harmonic (SH) of the 3.2~$\mu$m fundamental POV beam in ZnO (a-cut, thickness $\approx270~\mu$m) and GaSe ($100~\mu$m thickness). To generate the SH, we swap the position of the focusing and collimating lenses and place the ZnO/GaSe crystal in between them. The transverse intensity distribution of the generated SH BGV beams in the ZnO crystal are shown in Figs.~\ref{Fig5}(a1)-(a3), whereas those for SH POV beams are shown in Figs.~\ref{Fig5}(b1)-(b3) for TC of the fundamental beams $l_0 = 1, -3$, and $-5$, respectively. The TC of SH beams are found to be $l_{2}=2$, $-6$, and $-10$ (see the Supplementary Material), which are exactly twice the TC of the corresponding fundamental beams. Moreover, the ring diameter values for SH POV beams are extracted to be $2r_0=$(1369(5), 1365(5), 1374(5)) $\mu$m for the same TC. From Figs.~\ref{Fig5}(a1)-(a3) and (b1)-(b3), it is clear that the radius of the maximum intensity of SH BGV beams increases with increasing values of the TC, whereas the radius of the maximum intensity of SH POV beams hardly change with TCs. Notice that the transverse intensity distribution of vortex beams with equal and opposite sign TC remain similar. Therefore, the TC $l_{0}=1,-3$, and -5 can be used in our study in a strictly growing sense. Furthermore, in Figs.~\ref{Fig5}(c1)-(c3) and (d1)-(d3), we show SH intensity distribution of BGV and POV beams generated in the GaSe crystal. We generate these harmonic vortices by driving the SH generation process with a fundamental beam carrying TC $l_0 = 3, 5$, and $6$, which results in TC $l_{2}=6$, $10$, and $12$, respectively (see the Supplementary Material for more details on the TC measurements). To obtain higher TC for the fundamental beam, we employ a technique that cascades different SPPs (two SPPs in our case), as reported in Ref.~\cite{wang2018}. 

Similar observations (as reported in the ZnO crystal) are also made in the context of the radius of the maximum intensity of SH BGV and POV beams for this case. For SH POV beams, the extracted ring diameter values are $2r_0=$(1484(5), 1487(5), 1485(5)). Figures~\ref{Fig5}(e) and (f) compare the obtained beam size of SH BGV and POV beams generated in ZnO and GaSe crystals. For SH POV beams, we obtain the characteristic TC-independent beam size distribution, which confirms the "perfect" nature of the harmonic vortices. For SH BGV beams, in contrast, we find a linear dependence of the harmonic vortices beam size with their TC. Therefore, this demonstrates that SH POV beams display a constant beam size irrespective of the TCs, whereas, the beam size shows a monotonic increase with the TC for SH BGV beams, a behavior similar to fundamental BGV beams. Furthermore, the TC of SH vortices ($l_{SH}$) scales linearly with the harmonic order and the TC of the fundamental beam i.e., $l_{SH}=2l_{0}$, which validates the OAM conservation in the SHG process, also discussed in Ref.~\cite{Li_2025}. We note that for a better comparison between SH POV and BGV beams in both the materials, we divide the POV beam radius by a factor of four.

\subsection*{High-order harmonic generation with fs POV beams}

\begin{figure}[ht!]
\centering 
\includegraphics[width=1\linewidth]{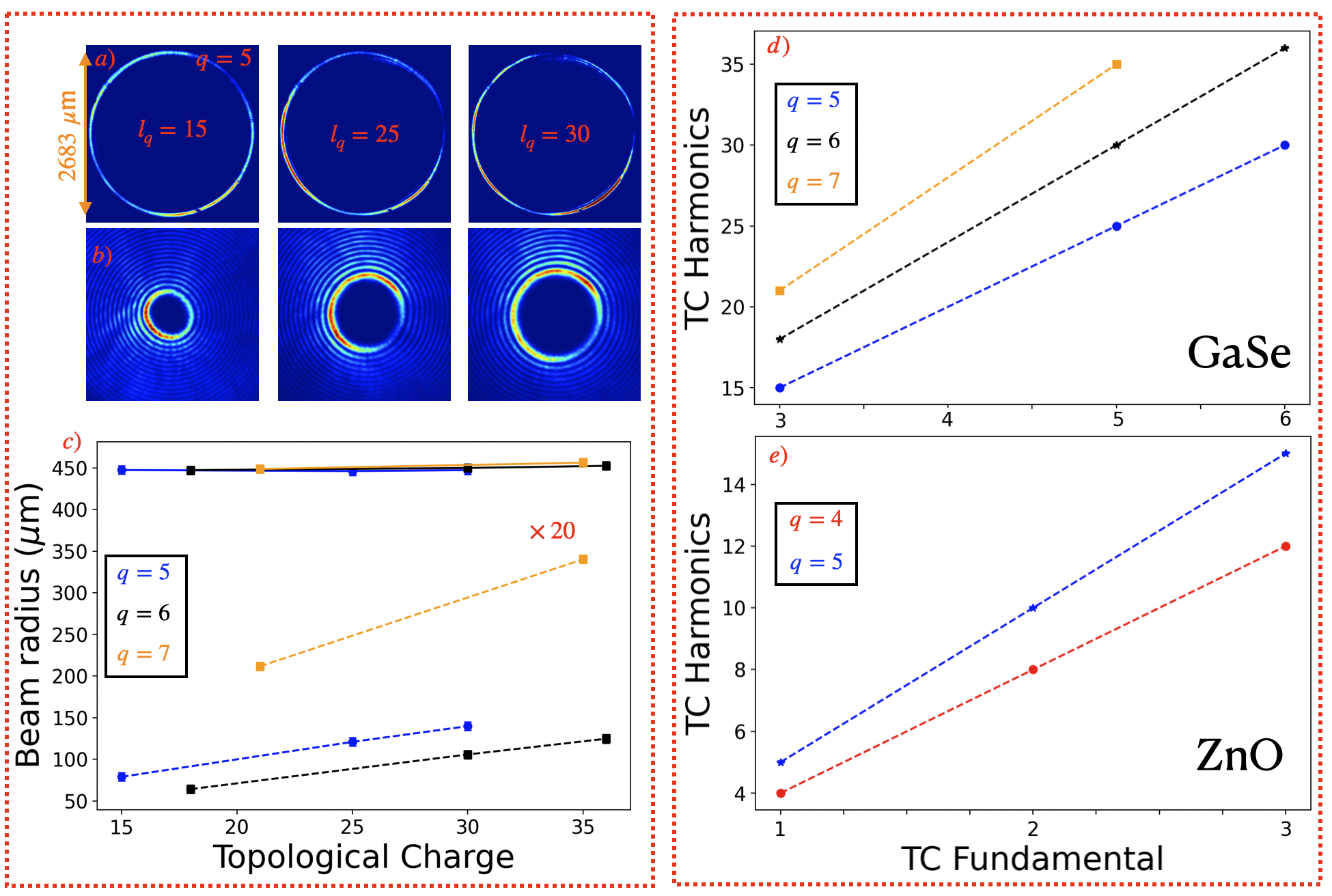}
\caption{Non-perturbative harmonics. (a) and (b), The $5^{\text{th}}$ POV and BGV harmonic beams, generated in GaSe, respectively. (c) Comparison between the beam size for different harmonic orders showing that the POV harmonic beams conserve similar beam radius values for a very large range of TCs. (d)-(e) Topological charge measurements in GaSe and ZnO, respectively. For the case of GaSe, we obtained harmonic vortices $q=6$ and 7, with TC as large as $l_q=35$ and 36, when the solid-state HHG process is driven by a fundamental field with $l_0=5$ and 6. There results clearly demonstrate that the up-conversion process can produce harmonics in the infrared-to-visible wavelength spectral range. Additionally, the linear trend followed by the harmonic TC as a function of the fundamental vortex beam TC, supports the conservation of OAM and extend its validity for this particular type of vortex beams. We divided the beam radius of the SH POV by 3 to make easily to compare the sizes of the vortex beams. The error bar for all the values corresponds to 5~$\mu$m.}
\label{Fig6}
\end{figure}

We now investigate non-perturbative high-order harmonic generation driven by femtosecond POV beams in ZnO and GaSe crystals, enabling us to assess the robustness of OAM transfer across materials with distinct dimensionality, band structure, and symmetry. The generation of high, non-perturbative POV harmonic beams also leverages on its conjugate nature with the BGV beams. After the AX element, BGV beams are transformed into POV beams at the lens focus. The fundamental and harmonic vortices then propagate in the ambient air and transform again into BGV beams. By using a strong focal lens ($f_c=75$~mm) after the target crystals (see Fig.~\ref{Fig1}), we convert the BGV harmonics into POV harmonics \cite{POV_FT}. For instance, in Fig.~\ref{Fig6}, we show the measured transverse intensity distribution for the $5^{\text{th}}$ POV and BGV harmonic beams generated from GaSe crystal for TC $l_{0}=3,5$ and $6$ of the fundamental beam (for harmonics generated from ZnO, see the Supplementary Material). As seen from Figs.~\ref{Fig6}(a)-(b), a single-ring intensity distribution is found for the POV harmonic beams, in line with the ring radius and ring width for different TCs, whereas high BGV harmonic beams show multi-ring intensity structures with enhanced radius of the first maximum intensity ring and dark core size for increasing values of the TC. In Fig.~\ref{Fig6}~(c), we compare the beam radius of different harmonic vortices, which are extracted from the experimental measurements: $2r_0=(2684(5), 2676(5), 2683(5))$, $2r_0=(2683(5), 2699(5), 2714(5))$, and $2r_0=(2691(5), 2736(5))$, for perfect harmonic vortices $q=5,6$ and 7, respectively. The above trend clearly depicts that the generated high-order harmonics indeed inherit the "perfectness" of the fundamental POV beams. Here, notice that we divide the beam diameter by a factor of six ($2r_{0}/6$) to plot the beam radius for different harmonic vortices generated by POV beams. In the case of BGV harmonic beams, the TC-dependent beam size is clearly evident from all the trends shown in Fig.~\ref{Fig6}~(c). Note that for the $7^{\text{th}}$ harmonic vortex, it was necessary to use a $\times$20 objective to properly resolve the beam in the focal position, which explains the high contrast of the beam size with the other two harmonic vortices. The extracted ring size values for BGV harmonic beams correspond to $2\rho_{BG}=(158(5), 242(5), 280(5))$, $2\rho_{BG}=(128(5), 212(5), 249(5))$, $2\rho_{BG}=423(4), 680(5)$. 

Finally, in Figs.~\ref{Fig6}~(d)-(e), we present the TC measurements for different harmonic vortices generated in GaSe and ZnO, which clearly follow a linear trend that scales with the TC of the fundamental vortex beams. As a result, the linear trend can be describe by a harmonic vortex OAM scaling law that follows $l_q=ql_0$, where $l_q$ is the TC of the harmonics. These results, support the OAM conservation in solid-state HHG and extend its validity for this particular type of structured light (see the Supplementary Material). The obtained OAM selection rule in solid-state HHG is identical to what we typically see in gas-phase HHG case, when a single driving OAM beam governs the harmonic generation process. Therefore, despite complex symmetries and geometries of different solid-state materials (unavailable in atomic gases), the OAM selection rule remains intact. Moreover, the generation of even harmonic vortices is allowed by the intrinsic symmetry of GaSe, which belongs to a non-centrosymmetric point group, and by the a-cut ($\langle 11\bar{2}0 \rangle$) ZnO crystal, where rotating the principal crystal axis by 90$^{\text{o}}$ with respect to the beam polarization effectively breaks the inversion symmetry of the light-matter interaction. 

Our results also demonstrate that it is possible to use the solid-state HHG process as a robust source of ultrashort POV and BGV beams with large TC in the IR-to-visible spectral range (see also the Supplementary material). This is explicitly show in Fig.~\ref{Fig6}, where TC as large as 36 are obtained by means of the OAM up-scaling in the non-linear process. To the best of our knowledge, these results have never been reported before.

\section*{Discussion}

Our experimental results demonstrate the generation of ultrashort MIR structured light and its parametric conversion into the IR and visible spectral regimes. They also represent a major advance in our quest to control light’s degrees of freedom and to explore different solid-state materials functionalities. In IR-to-visible spectral range, creating fs BGV and POV beams would require specialized laser systems to generate fs Gaussian beams and spatial light modulators to imprint the TC onto them. These light modulators are limited in efficiency, operation wavelength and bandwidth, all crucial for applications in nonlinear, non-perturbative physics in solids. We circumvent these limitations, offering a compact and flexible route to produce harmonic vortices with high values of the TC ($l_q=35$ for $q=7$) in the IR-to-visible spectral range. Furthermore, the HHG process driven by fs MIR vortex beams, make it is possible to investigate non-perturbative effects in ambient air. This result represents a major difference and advantage over the gas-phase studies which typically require ultrahigh vacuum conditions. 

Producing short wavelengths with large TC via HHG is fundamental to further expand the study of the helical phase effects in different physical systems. For example, understanding the absorption of molecular switches and motors dissolved in liquids is pivotal to increase the efficiency of energy conversion. Vortex beams provide a potential new route in this direction via the helical dichroism, since now it is possible to study if the molecular motors and switches are sensitive to the OAM sign and if new and more efficient OAM-dependent molecular transitions can be achieved. Most importantly, studying those systems via transient absorption spectroscopy, for example, allows for understanding the environment's effect on the molecular response to the helical light. In addition to this, the opportunity to generate and incorporate topological quantum light (quantum light carrying OAM) into ultrafast studies in solid systems could potentially open new opportunities to manipulate the harmonic spectrum basic characteristics. 

Beyond demonstrating control over the spatial structure of harmonics, our results confirm the conservation of OAM in solid-state HHG for the POV and BGV beams. The harmonic vortices TC scales linearly with both the fundamental OAM and the harmonic order, closely following the up-scaling law previously observed in gases and also recently demonstrated in several solid-state materials~\cite{POV_Fundamental}. This provides a robust experimental benchmark for theoretical models addressing microscopic and macroscopic electron dynamics in solids under helical strong-field excitation.

Generating harmonic vortices in semiconductor materials also expands the accessible spectral range of ultrashort structured light. Previously, harmonic vortex generation was largely restricted to the ultraviolet and extreme-ultraviolet in gases, where vacuum systems and low-efficiency XUV optics limit applications. Additionally, producing POV and BGV harmonic beams in the infrared-to-visible range opens opportunities for ultrafast transient absorption spectroscopy with OAM as a new degree of freedom and stimulated emission depletion microscopy, where the small ring size of the POV can be advantageous over previous approaches. It also enables studies of the fundamental optics of ultrashort structured light: we demonstrate the conjugate nature of POV and BGV beams during propagation (see Supplementary Material), opening the door to investigate propagation of ultrashort structured light in different media and greatly expanding their applicability in both fundamental and technological contexts. Additionally, with the advent of more complex topological structures, such as optical skyrmions and Hopfions, solid-state HHG provides an ideal platform to explore these exotic light fields and to produce them in the ultrashort regime.

To summarize, we demonstrated the generation of fs MIR BGV and POV beams. Our results confirm that POVs maintain even for fs pulses a TC-invariant beam size, whereas BGV beams always exhibit a TC-dependent beam size. Furthermore, these characteristics were shown to be preserved in the corresponding harmonic vortex beams, enabling the generation of POV and BGV beams in the shorter wavelength regimes with large TC. Most importantly, we demonstrate here the possibility to study nonlinear effects in solid-state materials with the use of structured light. Our work establishes a foundation for investigating the high-harmonic generation process driven by POV and BGV beams and the generation of attosecond twisted pulses in solid-state systems. 

\section*{Methods}

\subsection*{Fourier relationship between POV and BGV beams}

The experimental generation of POV beams relies solely on the optical FT of BGV beams. This relationship allows an arbitrary BGV beam to be mapped into a POV beam as follows \cite{POV_FT}: 

\begin{equation}
    \tilde{E}(r,\theta) = \frac{k}{2\pi i f}\int_0^\infty\int_0^{2\pi}E(\rho,\phi)e^{-\frac{ik}{f}\rho r cos(\theta-\phi)}d\rho \rho d\phi. \label{Integral}
\end{equation}

Here, $E(\rho,\phi)$ denotes the complex field amplitude of a BGV beam (given by Eq.\ref{BGV}), and $\tilde{E}(r,\theta)$ denotes the FT complex field amplitude of a POV beam, given by Eq.\ref{eq1}. Furthermore, $f$ is the focal length of the lens, $k=2\pi/\lambda$ represents the magnitude of the wave vector and $\lambda$ is the laser wavelength. Note that Eq.\ref{Integral} holds for both the fundamental BGV and POV beams, as well as for their harmonic counterparts (since BGV and POV harmonic beams are also Fourier pairs, as shown from our experimental results). Therefore, the helical phase of BGV harmonics, exp($iql\phi$), translates to the helical phase of POV harmonics, exp($iql\theta$). Accordingly, the TC of the POV beam can be extracted from the BGV beam, since it remains unchanged by the FT. This scenario corresponds exactly to the our experimental measurements aimed to extract the TC of the harmonic vortices.

\subsection*{TC measurement of BGV beams}
The TC measurement of the fundamental BGV and harmonic beams is based on the principle of the $\pi/2$ mode conversion. In this case, illuminating a CL with a BGV beam produces tilted HG beams surrounded by multiple rings in the detection plane. Quantitatively, this can be understood as follows: Under the small argument condition i.e., $k_r \rho \ll 1$, the Bessel function of the first kind of order $\ell$, describing the BGV beam, can be approximated as:    

\begin{equation}
    J_\ell(k_r \rho) \approx \frac{(k_r \rho)^\ell}{2^\ell \ell!}, \label{approximation}
\end{equation}

Here, $\ell$ represents the TC of the fundamental BGV and harmonic beams i.e., $\ell = l_0$ for the fundamental field and $\ell = ql_0$ for the harmonic field. Utilizing Eq.\ref{approximation}, we simplify Eq.\ref{BGV}, which now takes the form: 

\begin{equation}
    E(\rho,\phi) \approx  E_{0}'\,
\exp\!\left(-\frac{\rho^{2}}{w^{2}_0}\right) \frac{(k_r \rho)^\ell}{2^\ell \ell!} \exp(i\ell\phi) = E_1 \exp\!\left(-\frac{(x^{2}+y^2)}{w^{2}_0}\right) (x+iy)^\ell, 
\end{equation}

where $E_1$ condenses all the constant values. Note that the term, $\rho^\ell \exp(i\ell\phi)$, is replaced with $(x+iy)^\ell$, where $(x,y)$ denote the cartesian coordinates of the beam at the source plane ($z=0$) and are related to polar coordinates $(\rho,\phi)$ via: $x=\rho \cos{\phi}$ and $y=\rho \sin{\phi}$. Following the calculation presented in Ref.\cite{OAMtec}, we derive the complex field amplitude of the beam at a distance $z$ as: 

\begin{equation}
    E(x',y',z)\propto H_\ell(\alpha_1 x'+i \alpha_2 y'), \label{HGFinal}
\end{equation}

where, the parameters $\alpha_1$ and $\alpha_2$ depend on the beam waist and beam wavelength at a particular distance $z$. Furthermore, $(x',y')$ are the beam coordinates at a distance $z$. Equation~\ref{HGFinal} demonstrates the vortex beam transformation into a HG pattern, up to some additional prefactors. The interplay between those prefactors and the the Hermite polynomial determines the distance after the lens at which the input vortex beam transforms into tilted HG beams. This particular distance corresponds to the lens focal point. Most importantly, the number of intensity zeroes between the bright lobes, corresponding to the order of the Hermite polynomial, contains the TC information. We note that, to obtain the same transformed HG patterns as shown in Fig.\ref{Fig2}, it is necessary to use Eq.\ref{BGV} without any approximation. 
\\
\\
\section*{Acknowledgments}
The ELI-ALPS project (GINOP-2.3.6-15-2015-00001) is supported by the European Union and co-financed by the European Regional Development Fund. C.G. and B.K.D. acknowledged the support of the European laser Infrastructure ERIC through the program ``ELI Call for Users". C.R.G. acknowledged support from SECIHTI, México (CBF-2025-I-1804). We thank Andrew Forbes and Marcelo Ciappina for the valuable comments on the manuscript.

\bibliography{sample}





\section*{Author contributions statement}
C.G., D.R., B.K.D., B.K. and E.C. conceptualized the idea. C.G., D.R., R.S. and B.K. conducted the experiments. C.G., D.R. and R.S. analyzed the results. W.G., S.F., Q.Z. and C.G. supervised the project. C.G, C.R.G. and B.K.D. wrote the manuscript. All authors contributed equally to and approved the final version of the manuscript. 







\end{document}